\documentstyle[epsfig]{espart}%
\def\3{{\ss}}
\newcommand{\vect}[1]{{\bf #1}}
\def\op{{\bf D}}

\def\err{{\bf e}}
\def\rhs{{\bf f}}
\def\rest{{\bf v}}
\def\zm{{\bf w}}
\def\e{{\bf e}}
\def\f{{\bf f}}
\def\r{{\bf r}}
\def\S{{\bf S}}
\def\U{{\bf U}}
\def\Bn{{\bf B}_0}

\def\matr#1{{\bf #1}}
\def\bracket#1#2{{\langle #1 , #2\rangle}}
\def\Rset{{\bf R}}
\def\Cset{{\bf C}}
\newcommand{\calA}{{\cal A}}
\newcommand{\A}{\calA}

\newcommand{\calO}{{\cal O}}

\newcommand{\cuberes}{|_{[x]}}
\newcommand{\Laplace}{\Delta}
\def\dirac{{\,{\bf D}\!\!\!\! /\,}}
\def\tilde#1{\widetilde{#1}}

\def\bxi{{\xi}}
\def\bxitilde{{\tilde{{\bxi}}}}
\def\rho{\varrho}
\def\epsilon{\varepsilon}

\begin{document}
\begin{frontmatter}
\title{The Principle of Indirect Elimination}
\author{Martin B\"aker}
\address{II. Institut f\"ur Theoretische Physik
         der Universit\"at Hamburg,
         Luruper Chaussee 149,
         22761 Hamburg , Germany,
       e-mail : $<$baeker@x4u2.desy.de$>$}
\begin{abstract}
  The principle of indirect elimination states that an algorithm for
  solving discretized differential equations can be used to identify
  its own bad-converging modes. When the number of bad-converging
  modes of the algorithm is not too large, the modes thus identified
  can be used to strongly improve the convergence. The method
  presented here is applicable to any standard algorithm like
  Conjugate Gradient, relaxation or multigrid.  An example from
  theoretical physics, the Dirac equation in the presence of
  almost-zero modes arising from instantons, is studied. Using the
  principle, bad-converging modes are removed efficiently.  Applied
  locally, the principle is one of the main ingredients of the
  Iteratively Smooting Unigrid algorithm.
\end{abstract}
\end{frontmatter}

\section{Introduction}
Discretized differential equations lie at the heart of many simulation
algorithms in physics. A large variety of solution algorithms like
Conjugate Gradient, Overrelaxation, or Multigrid exist to deal
efficiently with such problems \cite{Press}. The convergence of these
algorithms
usually depends on the condition number of the problem operator,
i.e.~the quotient of its largest and smallest eigenvalue. (For many
simple problems multigrid methods will always converge well. Here we
are not interested in such cases.) When the number of eigenmodes with
very small eigenvalues is not large, each of these methods could be
accelerated if an additional method for dealing with these modes would
be applied.

In this paper we want to study a
method that can be used to do exactly this. It is partly based on the
multigrid idea and relies on a  surprisingly simple principle,
called the {\em Principle of indirect elimination\/} or {\em PIE}.

We will explain this principle in general context and then apply it to
a case where the occurence of almost-zero modes spoils the convergence
of standard methods, namely the Dirac equation in a gauge field
background with instantons \cite{Dilger,Sorin}. We will also show
connections to an idea by Kalkreuter somewhat similar in spirit,
called the {\em updating on the last point} \cite{ThomasII}, and
explain why our method is more general. Finally, we will briefly
remark on the connections to the Iteratively Smoothing Unigrid
algorithm \cite{ISU}.

\section{The general problem}
Consider a linear operator~$\op$ which may arise from a discretized
differential equation. Here and in the following we assume~$\op$ to
be {\em positive definite}, if it were not, we could use the
operator~$\op^\ast \op$ instead. The general form of the equation to
be solved is then
\begin{equation}
  \label{fundamental_eq}
  \op \bxi = \rhs \quad .
\end{equation}
Let us call the lowest eigenvalue of the operator~$\epsilon_0$
\footnote {It is not fully correct to speak about eigenvalues
  of~$\op$. In section~\ref{Dual} we will explain what is meant
  by such a statement.}.  Its value determines the {\em criticality\/}
of the operator because the smaller it is the larger the condition
number (quotient of largest and smallest eigenvalue)
of the operator will be. If $\epsilon_0=0$ the problem is
ill-posed because the contribution of this zero-mode to the solution
is not determined. For small $\epsilon_0$ standard iterative methods
will converge only slowly,
the
convergence time~$\tau$ (the number of iterations needed to reduce the
error by a factor of~e) behaving like $\tau\propto \kappa^{z/2}$,
where $\kappa$ is the condition number of $\op$ and $z$ is the
critical exponent. This behaviour is called {\em critical slowing down\/}
because the more critical the problem gets the slower the algorithm
will be. For relaxation methods, one usually finds
$z\approx2$, Conjugate Gradient has a critical exponent of
$z\approx1$. An optimal algorithm should have a critical exponent of~0.

At each time-step, any iterative method  will yield an approximate
solution~$\bxitilde$.
We introduce two important quantities: the {\em error\/}~$\e= \bxi
-\bxitilde$ which is the difference between the true and the actual
solution and is of course not known, and the {\em residual\/}~$\r= \f
- \op \bxitilde$, the difference between the true and the actual
righthandside. With these definitions we can recast the fundamental
equation~(\ref{fundamental_eq}) as
\begin{equation} \op \e = \r \quad, \label{error_eq} \end{equation}
called the {\em error equation}.

For a linear method, we can also introduce the {\em iteration matrix}~$\S$
which tells us what the new error after the next iteration step will
be, given the old one:
\begin{equation} \e^{\rm new} = \matr{S} \e^{\rm old}
  \quad. \label{IterError} \end{equation} The concrete structure of
  the iteration matrix is irrelevant for the following discussion, see
  \cite{Varga,Young} for examples. The important point here is that
  the iteration matrix is the reduction matrix for the error. Its
  eigenvalues should lie between minus one and one and convergence is
  governed by the eigenmode of~$\S$ with absolute value of the
  eigenvalue closest to one.

In the following sections we will usually assume the algorithm to be
linear because the existence of an iteration matrix eases the
analysis. Nevertheless the method presented here could be applied to
the Conjugate Gradient algorithm as well, see also section~\ref{instantons}
\subsection{Remark on vector spaces}
\label{Dual}
For the analysis it is important to distinguish between a vector space
and its dual \cite{Sokal}. The differential operator~$\op$ maps a
vector $\bxi\in V$ to a vector in the dual space $\rhs \in V^*$.
To see this, consider the Laplace equation in electrodynamics as an
example: $\Laplace \phi = - \rho$. The Laplace operator maps a
potential onto a charge density. These two objects can be regarded as
dual vectors because there is a unique way of assigning a real number
to them, namely the energy $\int \rho(x) \phi(x) dx$. The Laplace
operator therefore provides us with a bilinear form
$\bracket{\phi}{\psi}_\Laplace = \int \phi(x) (\Laplace \psi)(x)
dx$. However, there is no natural identification between the vector
space and its dual besides that given by this scalar product.
We will later see an example where one is easily drawn to wrong
conclusions if this distinction is not taken into account.

It is not really meaningful to speak about eigenvectors or -values of
bilinear forms. On the other hand, the iteration matrix of
relaxational methods maps the error to another error and is therefore
a map $\S : V \rightarrow V$, possessing eigenvectors. It are the
eigenvalues of this matrix that determine the convergence. The
standard identification of eigenvectors of $\S$ with eigenvectors of
$\op$ is done using additional structure. This is given by the matrix
$\Bn$ which is defined through the relation $\S = {\bf I} - \Bn^{-1}
\op$. (Standard relaxation methods arise from splitting the
fundamental operator $\op = \Bn + {\bf C}_0$, where $\Bn$ is chosen such
that it approximates $\op$ as good as possible but is ``easy to
invert''.) $\Bn$ is an additional bilinear form and furnishes us with
a scalar product in addition to the scalar product given by $\op$.

For Conjugate Gradient, the situation is similar: Conjugate Gradient
updating steps require computations of scalar products, e.g.\ $\alpha
= \bracket{\r}{\r} / \bracket{\bf d}{\op{\bf d}}$, where ${\bf d}$
is the search vector. Here we need another scalar product than the
$\op$-product.

It is therefore only correct to speak of eigenvectors of $\op$ when we
have chosen a basis that is in some sense natural. For example, if we
use the standard site-wise basis and find that the eigenvectors of
$\op$ in this basis agree with those of $\S$, the sloppy way of speach
is justified. This will be the case for the example we will
study below. Nevertheless, in the theoretical parts of this paper
we will be more strict.
\section{PIE in general}
After these preliminaries we formulate the
\begin{quote}
{\bf Principle of indirect elimination (PIE):\/} It is easier to {calculate\/}
the shape of a bad-converging mode for a certain algorithm than to
reduce it directly using this algorithm.
\end{quote}

To see this, consider the case where there is only one bad-converging
mode and all others are reduced efficiently by the algorithm. We now
use the algorithm to
{\em try\/} to solve an equation of which we already know the
solution, for example
the equation $\op \bxi =0$. In this case we have $\bxitilde = -\err$,
so we know the error as well. Remembering equation~(\ref{IterError})
we see that we can now directly investigate how the iteration matrix
acts. After~$n$ iterations we have
\begin{equation}\bxitilde^{(n)} = \S^n \bxitilde^{(0)} \quad , \end{equation}
  where $\bxitilde^{(0)}$ is the initial guess we started with and
  $\bxitilde^{(n)}$ is the approximate solution after the $n$-th
  iteration.  For $n\rightarrow\infty$ $\S^n$ projects onto the
  eigenvector of $\S$ with the largest absolute value of the
  eigenvalue, which  is the slowest-converging mode.  For
  finite $n$ the accuracy of the projection depends on quotient
  between the largest and the second-largest eigenvalue: The larger
  this is, the better the projection will be. (This can be seen easily
  by imagining $\S$ to be diagonalized.) In the model case considered
  here, where there is only one bad-converging mode, this quotient
  will be large and so $\S^n \bxitilde^{(0)}$ will converge rapidly
  against the bad-converging mode.

If the number of bad-converging modes is larger than one, but still
small, we can use the same technique to calculate them if we take
care of orthogonalizing the approximations to the already known
modes. By this it is obvious that this method will only be useful if
this number is not too large, otherwise the calculations will take too
much time. We will later comment on how the principle of indirect
elimination can be applied locally and used to construct a multigrid
algorithm.

Let us come back to the case of only one bad-converging mode. If we
have calculated this using the principle of indirect elimination, how
can we apply this knowledge to improve convergence?

The answer relies on multigrid ideas and is in fact very simple. Let
us call the bad-converging mode $\zm$. We define an operator $\A:
\Rset\rightarrow V, \mu \mapsto \mu \zm$ that creates a vector on the
fundamental lattice from a number.  This cumbersome notation has a
two-fold purpose: First it stresses the similarity to multigrid ideas,
where $\A$ would be called an interpolation operator, second it will
later allow us to study the case where $\A$ is not exactly equal to
the bad-converging mode $\zm$ to see how this will affect the
convergence.

To solve the inhomogenuous
equation, we first apply our standard iterative solver a few times.
This will reduce all components of the error appreciably except for a
part proportional to~$\zm$: $\err \approx c \zm$.
Inserting this knowledge into the error equation~(\ref{error_eq}) or
using the fact that $\r = \op \err \approx \op\zm c$ we get
\begin{equation}
\op (c \A) \approx \r \Longrightarrow \A^* \op \A c \approx \A^* \r \quad .
\end{equation}
In other words, we have transformed the fundamental equation, living
on a large lattice, into an equation for scalars (or simple matrices
in the case of a gauge theory, see below). This new equation can be
considered to live on a lattice with only {\em one point}. In
multigrid language this is often called the ``last-point lattice'' as
we have there a whole tower of coarser and coarser lattices of which
the last consists of only one point.

The equation on the last point can be solved easily to get~$c$ and
afterwards we correct our approximation: $\bxitilde \leftarrow
\bxitilde + \A c$.  Thus we have reduced that part of the error
corresponding to~$\A$. It is well-known from the multigrid context
that using the largest mode of~$\S$ as interpolation operator will
yield the best convergence (Greenbaum criterion \cite{Greenbaum}).  If
the iterative method used before has not been perfect, i.e.~if the
error still contains contributions from other modes, we now have to
start the iteration again to act on the remaining parts. This may
again introduce error-components proportional to $\A$ which are then
reduced by another ``last-point updating''.

We can now understand the reason why the principle has been called
principle of indirect elimination: Direct elimination of the
bad-converging mode using the iterative solver does not work
efficiently, but an indirect approach, first trying to solve an
auxilliary equation and only afterwards addressing the real problem,
works fine.

In practice the situation will not be the idealized one described
above. We now want to study two situations: What will the result of
the correction be when the error is not an exact
multiple of the zero-mode~$\zm$, and what happens when $\A$ deviates
{} from $\zm$?

In the first case it is easy to prove that
after the
correction $\bxitilde$ and $\A$ will be $\op$-orthogonal even if the
error before was not a multiple of $\A$, but contained an additional
contribution $\rest$:
\[
  -\bxitilde = \err = c \A + \rest \]
\[   \r = \op \err = \op \A c + \op \rest\]
  {\hbox{The equation on the last point is then:}}
\[   \A^* \op \A\, x = \A^* \op \A\,  c + \A^* \op \rest\]
\[   \Longrightarrow \qquad x = c +
   \frac{\displaystyle\A^* \op \rest}{\displaystyle\A^* \op \A}\]
  {\hbox{Correcting $\bxitilde$ yields}}
\[   \bxitilde
  =  \frac{\displaystyle\A^* \op \rest}{\displaystyle\A^* \op \A}\, \A
  - \rest\]
  {\hbox{This gives $\op$-orthogonality: }}
  \[ \A^* \op \bxitilde = + \frac{\displaystyle\A^* \op
    \rest}{\displaystyle\A^* \op \A}\, \A^* \op \A - \A^* \op \rest =
  0 \]
Now we want to investigate the second question, namely how well the
approximation of the zero-mode has to be.  To do so we can prove the
following rather trivial

\begin{thm}
We have an algorithm consisting of two parts. The
  first part is able to eliminate completely all components of the
  error except one single mode~$\zm$, so we have $\err=\zm$. The
  second updating then consists of an updating on the last point as
  described above using an approximation $\A$ of $\zm$.

We can split the bad-converging mode~$\zm$ into two $\op$-orthogonal
parts:
\begin{equation} \zm = \calA c + \rest \quad, {\rm \ with\ }
  \langle\calA ,\op\rest\rangle=0 \quad.\end{equation}
Then the iteration matrix ${\bf M}$ of the full algorithm consisting
of both steps has the (squared) energy norm (with respect to~$\op$)
\begin{equation} \|{\bf M}\|_\op^2 =
  \frac{\langle\rest,\op\rest\rangle}{\langle\zm,\op\zm\rangle}
\quad .\end{equation}
\end{thm}

\begin{pf}
The energy norm is defined as
\[ \|{\bf M}\|_\op^2 =
  \sup_\bxi \frac{\langle{\bf M}\bxi, \op{\bf
      M}\bxi\rangle}{\langle\bxi,\op\bxi\rangle} \quad.\]
Let $\S$ be the iteration matrix of the first part of the algorithm.
As it eliminates all parts of an arbitrary error except the
mode~$\zm$, it is clear that the supremum in the definition will be
reached for~$\bxi=\zm$. ${\bf S}$ does not affect~$\zm$.
After the iteration only that part of $\zm$ that is $\op$-orthogonal
to $\A$ will remain, see the calculation above.
So we get ${\bf M} \zm = \rest$.

Thus we have
\[ \|{\bf M}\|_\op^2 =
\frac{\langle{\bf M}\zm, \op{\bf
    M}\zm\rangle}{\langle\zm,\op\zm\rangle} =
\frac{\langle\rest,\op\rest\rangle}{\langle\zm,\op\zm\rangle}
\quad.\qquad\hfill\qed
\]

\end{pf}

It is also useful to look at this geometrically:
The
angle~$\theta$ between
the vector~$\zm$ and $\calA c$ with respect to the scalar product
defined by~$\op$ is given by
\[ \cos\theta =
  \frac{\langle\zm,\op\calA c\rangle}{\langle
    \zm,\op\zm\rangle^{1/2} \langle\calA c
    ,\op\calA c\rangle^{1/2}}\quad. \]
The
reduction works by first projecting $\zm$ onto the direction given
by $\calA$ and then taking the $\op$-orthogonal part of this.  This
orthogonal part is the vector~$\rest$; it is all that remains after
the coarse-grid correction step. The length of this vector is given
by $\|\rest\|_\op = \|\zm\|_\op \sin \theta$. The reduction factor,
which is equal to the norm of the iteration matrix,
is~$\sin\theta$:

\begin{center}
\epsfig{file=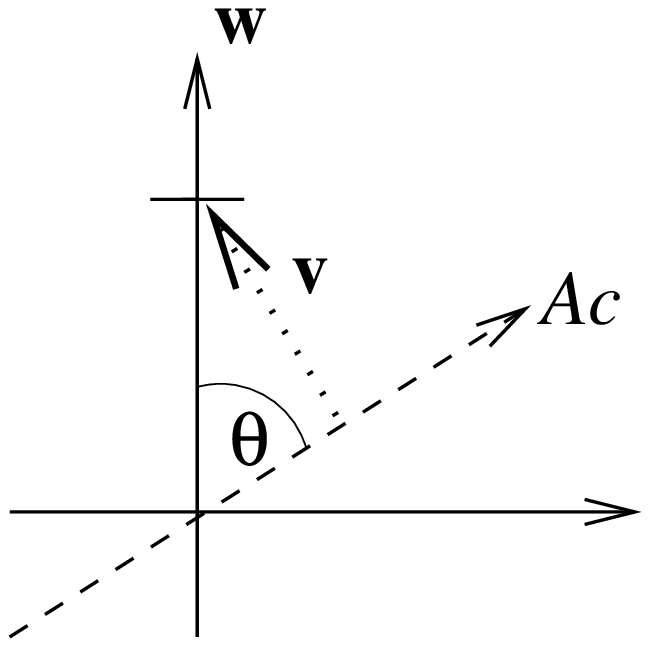,width=5cm}
\end{center}

Using Pythagoras' theorem we get
\[
  \sin^2\theta = 1-\cos^2 \theta = 1 -
  \frac{\langle\zm,\op\calA c\rangle^2}{\langle
    \zm,\op\zm\rangle \langle\calA c
    ,\op\calA c\rangle}\quad,\]
and inserting the split of the vector $\zm$ and again using the
orthogonality property, we finally arrive at
\[
\sin^2\theta = 1 - \displaystyle\frac{  (\langle\calA c,
  \op\calA c\rangle
 + \langle\rest,\op\calA c\rangle)^2}{\langle
    \zm,\op\zm\rangle \langle\calA c
    ,\op\calA c\rangle}
= \displaystyle\frac{\langle\zm,\op\zm\rangle-
  \langle\calA c,\op\calA c\rangle}{
  \langle\zm,\op\zm\rangle}
= \displaystyle\frac{\langle\rest,\op\rest\rangle}{\langle\zm,\op\zm\rangle}
\quad.\]

What are the implications of this theorem? First it must be understood
that the energy norm of the iteration matrix will be equal to the
spectral radius provided the matrix~${\bf S}$ is $\op$-symmetric,
i.e.\ ${\bf S}^* \op = \op {\bf S}$ \footnote{Actually, this is a nice
  example for the necesssity to distinguish endomorphisms and bilinear
  forms: treating $\op$ as an endomorphism, not as a bilinear form, we
  would transform it using the wrong relation and {\em loose\/} the
  $\op$-symmetry property after the transformation: We have $\S^* \op
  = \op \S$. Now if we transform both using the transformation for
  endomorphisms, we get $ (\S^\prime)^* \op^\prime = \left(\U^* \S^*
  (\U^{-1})^*\right) \U^{-1} \op \U$! Here there is no cancellation as it
  would be with the correct transformation: $ \S^{T^\prime} \op^\prime
  = \left(\U^* \S^* (\U^{-1})^*\right) \U^* \op \U$. Only if $\U$ is
  orthogonal or
  unitary do we get the same cancellation.}. This is true when the
operator~$\op$ does not mix the mode~$\zm$ with the other modes. We
can then regard $\zm$ as an eigenmode of $\op$ because the
matrix~${\bf S}$ provides us with an identification of the mode $\zm$
with the corresponding mode in the dual space.  Hence the energy norm
directly tells us about the convergence rate of the algorithm.  $\S$
will also be $\op$-symmetric for standard iterative methods
like Jacobi-relaxation and can be made so as well for SOR or
Gau\3-Seidel relaxation.

By choosing~$\zm$ as the zero-mode the theorem shows how important
the correct treatment of this mode is. The closer the range of the
interpolation operator $\calA$ is to the zero-mode and the smaller the
energy norm of the residual part $\rest$ the better the convergence
will be. (In the limiting case where the two are identical, the
difference vector is zero and the error of the zero-mode is eliminated
perfectly, as expected.) It is not only important that the
zero-mode is approximated well by the interpolation operator on the
last point, the convergence will also be better when the
difference vector between zero-mode and the mode used for the
interpolation has a small energy norm and is as smooth as possible.

The theorem  also serves to explain a finding by Kalkreuter
\cite{ThomasII}. He found that it is possible to eliminate critical
slowing down in a multigrid algorithm (actually a two-grid) for the
standard Laplace equation even with interpolation operators that are
not able to represent the zero-mode (which is a constant in this case)
exactly, but only approximately. In the light of our theorem this
could be understood if the difference vector has a small energy norm.
This, however, has not been tested.

Thus we have seen the importance of the correct treatment of the
zero-mode.  Other methods to remove convergence problems caused by the
zero-mode can be thought of.  Kalkreuter \cite{ThomasII} proposed a
simple rescaling of the approximate solution~$\bxitilde$ in addition
to a multigrid or a relaxation algorithm to improve the convergence.
This method completely eliminates critical slowing down in the
simplest model problem, the Laplace equation on a two-point grid.
The rescaling amounts to using the approximate solution $\bxitilde$
itself as interpolation operator~$\calA$.

The motivation for this updating scheme can be found in the following
argument: Consider again the equation
$\op \bxi =
{\bf f}$. Solving this gives $\bxi = \op^{-1} \vect{f} = (\Bn^{-1}
\op)^{-1} \Bn^{-1}f $ where we have inserted a unit matrix.
Let us fix the righthandside and increase the criticality of the problem.
The more critical it gets the smaller the lowest eigenvalue
$\epsilon_0$ of $(\Bn^{-1} \op)$ will be.
The solution $\bxi$ will then have larger and larger contributions
{} from the lowest eigenmode of $(\Bn^{-1} \op)$. Therefore $\bxi$ itself
will be a good approximation to the bad-converging mode and can be
used as interpolation operator.

We see that this method is very similar to
the principle of indirect elimination. However, the principle of
indirect elimination will always provide us with an approximation to
the zero-mode without any contribution from a given righthandside,
whereas Kalkreuter's method will only work well for large
criticality: The interpolation operators used by the methods are $\A =
\S^n \bxi^{(0)}$ for the principle of indirect elimination and $\A = \S^n
\bxi^{(0)} + \Bn^{-n} \rhs$ for Kalkreuter's method.
Even more important, the principle of indirect
elimination can be used several times to remove more than just one
mode, this is impossible with the other approach. On the other hand for
large criticality and the case of only one bad-converging mode,
Kalkreuter's method has the advantage of not needing auxilliary
iterations to calculate $\A$ because $\bxitilde$ is used.

Kalkreuter used this method in addition to a usual multigrid
or relaxation method. He found that there is no strong improvement for
a multigrid algorithm, but for standard local relaxation the
asymptotical critical slowing down (i.e.~critical slowing down for
fixed grid size and infinitely many iterations) was eliminated for the
Laplace equation with periodic boundary conditions. This is what we
expect for a method that treats the lowest mode of the problem correctly
because in this case it is the eigenvalue of the second-lowest mode
that determines the convergence and this scales with the size of the
grid, not with the lowest eigenvalue.  So for increasing grid
sizes critical slowing down should still be present; this in agreement
with Kalkreuter's results.
\section{Killing Instantons}
\subsection{The Dirac operator}
Our example for a discretized differential equation with a small
number of bad-converging modes is taken from theoretical high-energy
physics, namely the two-dimensional Dirac equation on the lattice in a
gauge field background with periodic boundary conditions.

For an introduction to Lattice Gauge Theory consult \cite{Creutz}.
Here we only present the framework:
Consider a regular, $d$-dimensional (hyper-)cubic lattice $\Lambda^0$
with lattice constant~$a$, lattice points~$z$ and directed
links~$(z,\mu)$.  The opposite link is then denoted by $(z+\mu,-\mu)$,
where $z+\mu$ means the next neighbour of $z$ in $\mu$-direction. The
direction index~$\mu$ runs from $-d$ to $d$. Usually, the lattices
used will be finite with an extension of $L$ points in each dimension
so that the number of degrees of freedom is~$n=L^d$.

A lattice gauge theory is defined by a gauge group~$G$ which might for
example be U(1) or SU(2). Elements of the gauge group act on a vector
space~$V$ which for the examples above would be $\Cset$ and $\Cset^2$,
respectively. The computations presented below were done in two
dimensions with gauge group U(1), so that instantons can occur. A
lattice gauge field (in this case) assigns a U(1)-``matrix''
$U(z,\mu)$ (which is simply a phase) to every link of the lattice,
subject to the condition $U(z,\mu) = U(z+\mu,-\mu)^{-1}$. These
matrices are distributed randomly with a Boltzmannian probability
distribution $\propto \exp (-\beta S_W(U))$, where $S_W$ is the
standard Wilson action of lattice gauge theory
\[S_W(U)= \sum_p Tr (1-U(\partial p)) \quad {\rm with}\ U(\partial p)
= U(b_4)U(b_3)U(b_2)U(b_1)\] for a plaquette~$p$ of the lattice with
links $b_1... b_4$ at its boundary. This distribution leads to a
correlation between the gauge field matrices with finite correlation
length $\chi$ for finite $\beta$. The case $\beta = 0$ corresponds to
a completely random choice of the matrices ($\chi=0$), for
$\beta=\infty$ all matrices are $\matr{1}$ ($\chi=\infty$). In this
sense, $\beta$ is a disorder parameter, the smaller $\beta$ the
shorter the correlation length and the larger the disorder.

The Dirac operator acts on matter fields~$\bxi$ living on the nodes of
the lattice. In Kogut-Susskind formulation \cite{Staggered} it is
defined as
\[
\left( \dirac \bxi\right)(z) =
\frac1a \sum_{\mu=1}^d \eta_{\mu, z} \left(U(z,\mu)^*\,\xi(z+\mu) -
U(z,-\mu)^* \,\xi(z-\mu) \right)
\quad.\]
Here the~$\eta_{\mu,z}= \pm 1$
are the remnants of the Dirac matrices~$\gamma^\mu$ in the continuum.

As the Dirac operator itself is not positive definite, we will use its
square~$\dirac^2$ in the following.  The squared Dirac has the
property of totally decoupling the even and odd parts of the lattice;
if we color the lattice points in checkerboard fashion, any red point
is only coupled to other red points, so that we can restrict our
attention to one of the sub-lattices. This will be especially useful
because it lifts the degeneracy of the eigenvalues: Usually each
eigenvalue of the Dirac operator is degenerated twice, but we can
choose the eigenvectors to live separated on the sub-lattices.  For
the sake of brevity we will generally speak of the Dirac operator even
when we
mean the squared Dirac.
\subsection{The instanton problem}
\label{instantons}
\begin{figure} \begin{center}
\epsfig{file=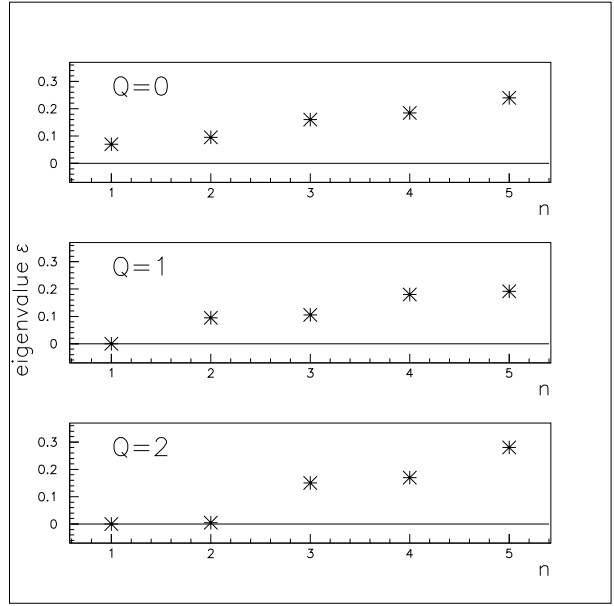,width=8cm}
\end{center}
\caption[The Atiyah-Singer theorem on the lattice]{The Atiyah-Singer
  theorem on the lattice: Shown are the five
  lowest eigenvalues of the staggered squared Dirac operator in a
  U(1) gauge  field at~$\beta=10$ on an~$18^2$-lattice for different
  topological charges~$Q$. Only one of the two sub-lattices is taken
  into account, on the full lattice each eigenvalue is degenerated twice.
\label{Qtopspectrum}}
\end{figure}

Many algorithms for solving the Dirac equation become problematic in
the presence of instantons. Instantons are gauge field configurations
that are topologically non-trivial but possess zero energy. Such
configurations are only possible for certain choices of the dimension
and the gauge group, in two dimensions instantons can occur when the
gauge group is U(1), see \cite{Nakahara} for an introduction.
The Atiyah-Singer theorem states that at
instanton charge~$Q$
the spectrum in the continuum will possess $2|Q|$ exact
zero-modes~\cite{Atiyah}; these become modes with extremely small
eigenvalues on the lattice \cite{Dilger}. For the purpose of this
paper it is not necessary to have an understanding of what an
instanton is, it is only important that they are special gauge field
configurations giving rise to almost-zero-modes on the lattice.
Figure~\ref{Qtopspectrum} shows the lower part of the spectrum of the
squared Dirac operator on an $18^2$-lattice at $\beta=10$ for
different instanton charges~$Q$, taking only one of the two
sublattices into account. Clearly the Atiyah-Singer theorem is
nicely reflected on the lattice.

When instantons are present,
the condition number of the Dirac operator becomes very large and the
problem is ill-posed. In \cite{Sorin} this problem is investigated in
detail for the Parallel Transported Multigrid.  In this section, we
want to use the principle of indirect elimination to show how an
algorithm which converges well in the absence of instantons can be
adapted to a case with instantons.

The idea is very simple: If~$m$ bad-converging modes are present, use
the principle of indirect elimination $m$ times to calculate
approximations $\A^i$ to these modes.

The method presented here could be applied to a Conjugate Gradient
algorithm. For this algorithm, Dilger \cite{Dilger} has found
that the number of iterations needed strongly increasses with the
instanton charge, therefore Conjugate Gradient would benefit from
the application of the method described here.

However, we will choose the ISU algorithm on small lattices and at
quite large values of~$\beta$ for a U(1) gauge field as an example.
This algorithm has been described in detail in~\cite{ISU}. As it is in
some parts based on the principle of indirect elimination, some
remarks will be made on this method in the next section. For this
section it is not necessary to understand how ISU works, it suffices
to know that for the parameters chosen the ISU algorithm converges
well for instanton charges 0 or $\pm1$ at large $\beta$ but badly for
larger instanton charges. The reason is that the algorithm in its
standard form contains one interpolation operator on the last point
(which is calculated as an approximation to the zero-mode) and so it
is able to eliminate one zero-mode, but not more.

In the improved algorithm
one tries to solve the equation $\op \A^i=0$ with the given algorithm.
As it eliminates all other modes quickly the approximate
solution will converge against a linear combination of the
bad-converging modes. Then we start the procedure again, but now
orthogonalizing the approximate solution to the interpolation operator
we already know, doing this successively for all~$m$ bad-converging
modes. (As the instanton charge can be easily measured, one usually
knows beforehand how many operators are needed; if one does not for
some reason, a dynamical approach can be chosen: Simply proceed
calculating the next interpolation operator until the convergence rate
of the trivial equation becomes good enough.)

The overall work for this procedure is proportional to the square of
the number of bad-converging modes, as is the work of actually
applying the interpolation operators to eliminate them. (The number
gets squared because of the need to calculate an effective
operator on the last point layer. However, the effective operator only
needs to be calculated once for each configuration.) This restricts
the method to cases where the number of bad-converging modes is not
too large, which usually is the case for instanton charges.

\begin{figure} \begin{center}
\epsfig{figure=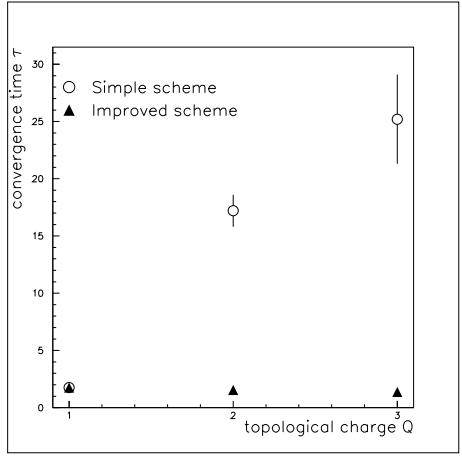,width=6cm}
\end{center}
\caption[Performance of standard and improved ISU for instanton
problems]{Performance of the standard and the improved ISU algorithm
  for the elimination of instanton modes. The data were generated on
  one sub-lattice of
  an~$18^2$-lattice at $\beta=10$ with a U(1) gauge field. The
  improved algorithm uses $Q$~interpolation operators on the last
  point to eliminate the almost-zero modes. (The number of
  configurations evaluated for the different topological charges
  was~217, 113, and 22. )
\label{KillingPic}}
\end{figure}

Figure~\ref{KillingPic} shows the performance of the usual ISU method
compared to the improved version for the Dirac operator in a U(1)
gauge field with different instanton charges.  We measured the
asymptotic convergence time, i.e.~the number of iterations
asymptotically needed to reduce the error by a factor of~e.  The
improved version of ISU used a number of interpolation operators on
the last point equal to the instanton charge which equals the number
of bad-converging modes. The data were generated on one sub-lattice of
an $18^2$-lattice at $\beta=10$. For this high value of $\beta$, the
convergence in the absence of instantons is good, as can be seen from
the value at $Q=1$.  The standard method works well for instanton
charge~0 or 1, as explained above, and its sensitivity to higher
instanton charges is striking.  The improved method shows no
dependence on the instanton charge, the convergence is good in all
cases. Note also that the standard deviation is much higher for the
usual method because it is affected by fluctuations in the eigenvalues
of the bad-converging modes.

Clearly the improved method is superior---the cost of
calculating the instanton modes is about 10~iterations for each
instanton plus the cost of the orthogonalization, whereas the saving
in the solution of the final equations is of the order of hundreds of
iterations depending on how much we want to reduce the residual.
\section{PIE and ISU}
We have seen above that the principle of indirect elimination will
only be helpful when the number of bad-converging modes is not too
large. In the case of simple relaxation methods, however, this number
is of order $\calO(n)$, so
storing them would cost ${\cal O}(n^2)$, where $n$ is
the number of degrees of freedom in the system. So it seems that the
method is useless in such cases.
In this section we want to explain how the Iteratively Smoothing
Unigrid algorithm (ISU) presented in
\cite{ISU} can be regarded as the local application of the principle
of indirect elimination. We will only present the basic idea here.
Some familiarity with the basic multigrid idea is assumed in this
section, see \cite{Brandt,Briggs,Hack} for introductions.

We can associate a length scale (e.g.\ a wavelength) with each modes
of our system.  Because they are local, relaxation methods eliminate
all those modes corresponding to a length scale of the order of one
lattice spacing. Usually there will be $\calO(n/2)$ of these.  Of the
remaining modes $\calO(n/4)$ will be associated with length scale $2a$
($a$ is the lattice spacing), $\calO(n/8)$ with scale $4a$ and so on.
So there will be many bad-converging modes with a small length scale
and only a few corresponding to a large scale. The ISU algorithm is a
method to calculate interpolation operators that are able to span the
space of these modes. These operators are restricted to parts of the
lattice, the size of the domain being determined by the scale of the
mode that is to be approximated by the operator.

To be more specific, let us start with the smallest scale $2a$. As in
usual multigrid methods, we divide the hypercubic lattice into
(overlapping) hypercubes or blocks $[x]$ of side length~3. Then we try
to solve the equation $\op\cuberes \calA_x^{[1]}(z) =0$ using a
relaxation method. Here $\op\cuberes$ is the restriction of $\op$ to
the block $[x]$ using Dirichlet boundary conditions.  What remains
after a few iterations will be the slowest-converging mode on this
scale and can therefore be used as interpolation operator on the first
block-lattice.  Repeating this for all the small hypercubes, we know
the shapes of the bad-converging modes on scale $2a$. Now we do the
same on the next scale, dividing the lattice into larger blocks (of
side length~7, agreeing with the formula $2^j-1$). Again we try to
solve the equation $\op \cuberes \calA_x^{[2]}(z) = 0$, where $[x]$
now denotes the larger blocks. The important point is that we use the
interpolation operators on the smaller scale that are already known
for this calculation to eliminate contributions from the
bad-converging modes on the smaller scale. In this way we proceed to
larger and larger hypercubes, always using the interpolation operators
already known.  This method would only fail if a large number of
bad-converging modes lived on a large length-scale.

It has been found that this algorithm is able to eliminate critical
slowing down completely for the case of the two-dimensional Laplace
equation in an SU(2) gauge field background at arbitrarily large
values of the gauge field disorder and the lattice size. An improved
version has been shown to do the same for the two-dimensional squared
Dirac equation, except for extremely large disorder ($\beta \approx
2$ or smaller).  See \cite{ISU} for details.

An idea that is similar in spirit to the principle of indirect
elimination has been discussed in \cite[section~4.6]{Achi}.  Brandt
proposes to do relaxations on the fundamental lattice with arbitrary
starting vectors to determine ``typical shapes of a slow-to-converge
error'' which could then be used to determine good multigrid
interpolation operators. Unfortunately, this idea suffers from a
severe disease: The number of modes that converge badly under simple
relaxation is huge (about half of the number of grid points). What one
will get by this procedure is a mixture of low-lying eigenmodes with
contributions depending on their eigenvalues. The time needed to
arrive at a function that consists only of the lowest eigenmodes will
be proportional to the lattice size, so the method will not work
without critical slowing down.

The difference to the ISU algorithm is that this is a so-called {\em
  unigrid} method. It allows for interpolation operators living on
different length-scales, whereas standard multigrid algorithms only
use interpolation operators living on small domains.  On each length
scale we need not represent all modes that converge badly on this and
on all higher scales; only the modes that belong to the scale
corresponding to a certain lattice constant have to be dealt with. The
next-coarser length-scale will then take care of the modes
corresponding to this scale and in their computation the smaller
scales are already taken into account.
\section{Conclusions}
We have presented a simple method to improve the convergence of
solution algorithms for discretized differential equations when the
number of bad-converging modes is small. The principle of indirect
elimination used to do this is based on the general idea that an
algorithm can be used to identify its own bad-converging modes.
Conceptionally, the method is similar to the general idea of
accelerating algorithms described in \cite{Sorin2}: One tries to find
out what the slow modes of the algorithm are and uses this knowledge
to improve the algorithm. For example, multigrid methods are based on
the fact that the slow modes are the smooth modes that can be obtained
by smooth interpolation. The principle of indirect elimination serves
to automatize this process in the case when the number of slow modes
is small so that it suffices to know them without doing further
analysis of their structure.

The
method has been studied for the case of the Dirac equation in a gauge
field background with instantons and worked extremely well. Applying
it locally leads to a multigrid method called the Iteratively
Smoothing
Unigrid.
\begin{ack}
I wish to thank Gerhard Mack and Alan Sokal for stimulating
discussions. Hermann Dilger provided me with a copy of his program for
generating U(1) gauge field configurations.
Financial support by the Deutsche Forschungsgemeinschaft
is gratefully acknowledged.
\end{ack}

\end{document}